\documentclass[useAMS,usenatbib,onecolumn,letterpaper]{mn2e}
\usepackage{graphicx}
\usepackage[
colorlinks=true,
	linkcolor=blue]{hyperref}

\newcommand{\pcm}{\,cm$^{-3}$}
\newcommand{\bfm}[1]{\mbox{\boldmath $#1$}}

\newcommand{\mat}[1]{\mbox{\boldmath $\mathsf{#1}$}}

\title[Driven multifluid MHD turbulence]{Driven
	Multifluid MHD Molecular Cloud Turbulence}
\author[T.P. Downes]{T.P. Downes$^{1,2,3}$\thanks{E-mail: turlough.downes@dcu.ie (TPD)}\\
$^{1}$School of Mathematical Sciences, Dublin City University,
	      Glasnevin, Dublin 9, Ireland\\
$^{2}$School of Cosmic Physics, Dublin Institute for Advanced Studies, 31 
Fitzwilliam Place, Dublin 2, Ireland \\
$^{3}$National Centre for Plasma Science and Technology, Dublin City 
University, Glasnevin, Dublin 9, Ireland}

\begin{document}

\date{Accepted 2012 June 22. Received 2012 June 21; in original form 2012 February 10}

\pagerange{\pageref{firstpage}--\pageref{lastpage}} \pubyear{2012}

\maketitle

\label{firstpage}

\begin{abstract}
It is believed that turbulence may have a significant impact on star formation and the dynamics and
evolution of the molecular clouds in which this occurs.  It is also known that non-ideal
magnetohydrodynamic effects influence the nature of this turbulence.  We present the results of a 
numerical study of 4-fluid MHD turbulence in which the dynamics of electrons, ions, charged dust
grains and neutrals and their interactions are followed.  The parameters describing the fluid being
simulated are based directly on observations of molecular clouds.  We find that the velocity and 
magnetic field power spectra are strongly influenced by multifluid effects on length-scales at least as 
large as 0.05\,pc.  The PDFs of the various species in the system are all found to be close to
log-normal, with charged species having a slightly less platykurtic (flattened) distribution than the 
neutrals.  We find that the introduction of multifluid effects does not significantly alter the 
structure functions of the centroid velocity increment.
\end{abstract}


\section{Introduction}

It is generally believed that molecular clouds are turbulent (see the reviews
of \citealt{mac04,elm04}).  Observationally \citep[e.g.][]{larson81, brunt10} this 
turbulence appears to be highly supersonic with RMS Mach numbers of anything up to 
20 or more.  It is also thought that the Alfv\'enic Mach number is not significantly less than one,
and may be much greater \citep[e.g.][]{heyer12}.  The amplitude of this turbulence makes it likely to 
be an important ingredient in both the dynamics of molecular clouds and the
process of star formation \citep{elm93,kle03} and hence developing an 
understanding of this phenomenon is of considerable interest.

Many authors have addressed the issue of MHD turbulence in the context
of molecular clouds using both the ideal MHD approximation
\citep{maclow98, maclow99, ost01, ves03, gus06, glover07, lem08, lem09, brunt10a, brunt10b,
	price11} and, 
more recently, various flavours of non-ideal MHD \citep{ois06, li08,
kud08, dos09, dos11}.  Three dimensional MHD turbulence involves the transfer 
of energy from an energy injection scale to ever smaller scales until the dissipation length scale of 
the system is reached. Given that in molecular clouds the lengthscales at which multifluid
effects become important are much larger than the viscous lengthscale, it is inevitable that
these effects will have an impact on the energy cascade. 

When ambipolar diffusion is included in MHD turbulence
simulations it has been found that it causes greater temporal variability 
in the turbulence statistics \citep{ois06, li08, kud08}.  Although clearly of 
lesser significance \citep{wardle99, dos09, dos11}, the Hall effect is capable 
of inducing topological changes in the magnetic field which are quite distinct 
to any influence caused by ambipolar diffusion.  In particular, it introduces a 
handedness into the flow which is of particular interest when considering the 
conversion of kinetic energy into magnetic energy.  Researchers working on 
reconnection and the solar wind have studied the Hall effect in the context of 
turbulence and found that, although the overall decay rate appears not to be 
affected, the usual coincidence of the magnetic and velocity fields seen in 
ideal MHD does not occur at small scales \citep{mat03, min06, ser07}.

\citet[hereafter Paper I]{dos09} performed simulations of decaying,
non-ideal MHD, molecular cloud turbulence incorporating parallel resistivity, 
the Hall effect and ambipolar diffusion.  They found that the Hall effect has
surprisingly little impact on the behaviour of the turbulence: it does
not affect the energy decay at all and has very limited impact on the
power spectra of any of the dynamical variables, with the exception of
the magnetic field at very short length-scales.  \citet[hereafter Paper
II]{dos11} extended the results of Paper I to properly multifluid MHD
turbulent decay.  A system of three fluids, 1 neutral and two charged
species, was simulated and, intriguingly, it appears that for the chosen
system a full multifluid simulation is unnecessary and simple non-ideal
MHD with temporally and spatially constant resistivities can give quite reliable results.

In this paper we use the {\sevensize HYDRA} code \citep{osd06, osd07} to 
investigate 
{\em driven}, isothermal, multifluid MHD turbulence in a system consisting of 
4 fluids (1 neutral and 3 charged).  The aim of this work is to determine the 
influence of multifluid effects on the behaviour of driven turbulence in 
molecular clouds.  An added benefit of our properly multifluid approach
is that we can self-consistently deduce the behaviour of the charged
species and, in principle, make links with observations such as those of
\citet{lh08}. 

The structure of this paper is as follows: in section
\ref{sec:model} we outline the equations and numerical
method; in section \ref{sec:initial-conditions} we specify our
parameters, initial conditions and how we drive the turbulence in our
simulations; section \ref{sec:resolution-study} contains the results of
a resolution study demonstrating that the results we present are
reasonably well converged; section \ref{sec:results} contains the
results and analysis of our simulations and finally we draw conclusions
from our work in section \ref{sec:conclusions}.  We note here that we omit 
a discussion of the differences in line widths between the charged and the neutral species.  The
results associated with these differences require detailed discussion in their own right and are
the subject of a forthcoming paper.

\section{The model}
\label{sec:model}

Molecular clouds are, to a great extent, weakly ionised systems.  This
allows us to make several simplifying assumptions, as follows:
\begin{itemize}
\item The bulk flow velocity {\em is} the neutral velocity
\item The majority of collisions experienced by each charged species
occur with neutrals
\item The charged species' inertia is unimportant
\item The charged species' pressure gradient is unimportant
\end{itemize}
Making these assumptions allows us to derive a relationship between the
electric field and the current density. This generalised Ohm's law
\citep[e.g.][]{falle03, cr02} allows us to avoid having to
calculate the electric field explicitly from the charge distribution.

\subsection{Equations}
\label{sec:equations}

Once we have our generalised Ohm's law, the system of equations for our
weakly ionised flow is as follows \citep[e.g.][]{falle03}

\begin{eqnarray}
\frac{\partial \rho_i}{\partial t} + \bfm{\nabla} \cdot \left(\rho_i
      \bfm{u}_i\right)  & = & 0 \mbox{, ($1 \leq i \leq N$), } \label{mass} \\
\frac{\partial \rho_1 \bfm{u}_1}{\partial t}
		   + \mathbf{\nabla}\cdot\left( \rho \bfm{u}_1 \bfm{u}_1 +
					a^2\rho\mat{I}\right) &
			   = & \bfm{J}\times\bfm{B} , \label{neutral_mom} \\
\frac{\partial \bfm{B}}{\partial t} + 
  \mathbf{\nabla}\cdot(\bfm{u}_1\bfm{B}-\bfm{B}\bfm{u}_1) & = & \mathbf{\nabla}\times\left\{
  r_0 \frac{(\mathbf{J}\cdot\mathbf{B})\mathbf{B}}{B^2} +  r_1 \frac{\mathbf{J} \times \mathbf{B}}{B} + r_2
	   \frac{\mathbf{B}\times(\mathbf{J}\times\mathbf{B})}{B^2}\right\}	, \label{B_eqn} \\
\alpha_i \rho_i\left(\bfm{E}' + \bfm{u}_i \times \bfm{B} \right) & = &
-  \rho_i \rho_1 K_{i\,1}(\bfm{u}_1-\bfm{u}_i) \mbox{, $2 \leq i \leq N$, }\label{charged_mom}\\
\mathbf{\nabla}\cdot\bfm{B} & = & 0 \label{divB} , \\
\mathbf{\nabla}\times\bfm{B} & = & \bfm{J} , \label{eqn-J} \\
\sum_{i=2}^N \alpha_i \rho_i  & =  & 0 . \label{charge_neutrality}
\end{eqnarray}
where $i$ denotes the species (with $i=1$ denoting the neutral species),
$\rho_i$, $\bfm{u}_i$, $a$, $\bfm{B}$, $\mat{I}$, $\bfm{J}$ and $\bfm{E}'$ 
are the mass density of species $i$, the velocity of species $i$, the 
isothermal sound speed, the magnetic field, the identity matrix, the current 
density and the electric field in the neutral rest frame, respectively.  The 
parameters $\alpha_i$ and $K_{i1}$ are the charge-to-mass ratios and the 
collision coefficient with the neutrals for species $i$.

These equations describe conservation of mass for all species, the
neutral momentum equation, the induction equation, force balance for the
charged species, the absence of magnetic monopoles, Faraday's Law and
charge neutrality.  In the work presented here we set $N=4$.

The electric field in the frame of the fluid, $\bfm{E}'$, is calculated
from the generalised Ohm's law for weakly ionised fluids (e.g.\  
\citealt{falle03, osd06}) and is given by
\begin{eqnarray}
\bfm{E}' = \bfm{E}_{\rm 0} +\bfm{E}_{\rm 1} +\bfm{E}_{\rm 2} , \label{eqn_E1}
\end{eqnarray}
where
\begin{eqnarray}
\bfm{E}_{\rm 0} & = & (\bfm{J}\cdot\bfm{a}_{\rm 0})\bfm{a}_{\rm 0} ,
		   \label{eqn_EO1} \\
\bfm{E}_{\rm 1} & = & \bfm{J}\times\bfm{a}_{\rm 1} , \label{eqn_EH1} \\
\bfm{E}_{\rm 2} & = & -(\bfm{J}\times\bfm{a}_{\rm 2})\times\bfm{a}_{\rm 2}
, \label{eqn_EA1} 
\end{eqnarray}
using the definitions $\bfm{a}_{\rm 0}\equiv f_{\rm 0} \bfm{B}$,
$\bfm{a}_{\rm 1}\equiv f_{\rm 1} \bfm{B}$, $\bfm{a}_{\rm 2}\equiv
f_{\rm 2} \bfm{B}$, where $f_{\rm 0}\equiv \sqrt{r_{\rm 0}}/B$,
$f_{\rm 1}\equiv r_{\rm 1}/B$, $f_{\rm 2}\equiv \sqrt{r_{\rm 2}}/B$. 
$r_{\rm 0}$, $r_{\rm 1}$ and $r_{\rm 2}$ are the parallel, Hall and
ambipolar resistivities respectively and are given by
\begin{eqnarray}
r_{\rm 0} & = & \frac{1}{\sigma_{\rm 0}} , \\
r_{\rm 1} & = & \frac{\sigma_{\rm 1}}{\sigma_{\rm 1}^2 + \sigma_{\rm 2}^2} , \\
r_{\rm 2} & = & \frac{\sigma_{\rm 2}}{\sigma_{\rm 1}^2 + \sigma_{\rm 2}^2} ,
\end{eqnarray}
\noindent with the conductivities given by
\begin{eqnarray}
\sigma_{\rm 0} & = & \frac{1}{B}\sum_{i=2}^N \alpha_i \rho_i \beta_i , \\
\sigma_{\rm 1} & = & \frac{1}{B}\sum_{i=2}^N \frac{\alpha_i \rho_i}
{1+\beta_i^2} ,\\
\sigma_{\rm 2} & = & \frac{1}{B}\sum_{i=2}^N \frac{\alpha_i \rho_i 
\beta_i}{1+\beta_i^2} ,
\end{eqnarray}
\noindent where $\beta_i$ is the Hall parameter for species~$i$ and is given by
\begin{eqnarray}
\label{eqn-hallpar}
\beta_i = \frac{\alpha_i B}{K_{1\,i}\rho_1} .
\end{eqnarray}

\subsection{Numerical method}
\label{sec:numerical-method}

Equations (\ref{mass}) -- (\ref{charge_neutrality}) are a nonlinear
system of PDEs and, in general, must be tackled using simulations.  With 
modern techniques the system is relatively straightforward, with the exception
of dealing with the right-hand side of equation (\ref{B_eqn}).  The
method used by {\sevensize HYDRA} to integrate these equations has been elucidated
previously in the literature \citep[e.g.][]{osd06, osd07} so we only give
a brief recapitulation of this here and refer the reader to these papers
for a more in-depth discussion.

The system is solved using an operator-split approach with three
operators:
\begin{itemize}
\item The equations governing the neutrals are solved using a standard second order, shock
capturing Godunov scheme.  The induction equation is also solved as part of 
this operator, but without incorporation of the non-ideal effects.  The divergence of the
	magnetic field is controlled using the method of \cite{dedner02}.
\item The magnetic diffusion terms are evaluated and applied using the
super-time-stepping method (for the ambipolar and parallel resistivity
terms) and the Hall Diffusion Scheme (for the Hall term). 
\item The velocities of the charged species are determined from equation
(\ref{charged_mom}) and the distribution of their densities are updated
using a second order upwind scheme.
\end{itemize}
The whole system of operators is stepped forward in time using Strang
operator splitting in order to ensure second order accuracy in time.

Since molecular cloud material has very low kinematic viscosity \cite[e.g.][]{jd12}, 
flows, these systems have a high Prandtl number and thus we can consider them to be effectively
inviscid.  As with all higher-order schemes, our simulations incorporate artificial viscosity
which smears discontinuities over around 3 -- 5 grid zones.  This artificial viscosity length
scale is much less than that over which the multifluid effects act (see, for example, figure
\ref{fig:v-spec-res}) and thus our multifluid effects dominate at all scales above around 3 -- 5
grid zones.  Formally, the Prandtl number of the highest resolution simulations in this work is around 
15.

The method is explicit and scales extremely well on massively parallel
systems, displaying strong scaling from 8192 up to 294,912 cores with 73\% 
efficiency on the JUGENE BlueGene/P system in J\"ulich, Germany.

\section{Numerical set-up}
\label{sec:initial-conditions}

We run our turbulence simulation in a cube of side 0.2\,pc with periodic
boundary conditions on all faces.  The density is initially uniform and
has the value of $10^4$\pcm, assuming that the average neutral particle
has a mass of 2.33\,$m_{\rm p}$ (where $m_{\rm p}$ is the mass of the
proton).  The magnetic field is also initially uniform of strength 20\,$\mu$G 
respectively.  The (isothermal) sound-speed is chosen to be 
$1.88\times10^4$\,cm\,s$^{-1}$ or, equivalently, the temperature is 10\,K.  The 
four fluids in the system are initially at rest.  These parameters were
chosen to be representative of molecular clouds on scales of 0.2\,pc and
are based on the data in Table 1 of \cite{cru99}.  The grid used is 
uniform and made up of $N_x \times N_y \times N_z$ grid points with 
$N_x = N_y = N_z = N_{\rm g}$.  We emphasise here that this work is directed 
at understanding the nature of multifluid MHD turbulence.  A physical system with the properties just
described will have a mass just below the Jeans mass and so may well be subject to dynamically 
important self-gravity. However, it seems reasonable that the optimal way of 
approaching understanding the behaviour of molecular clouds is to first 
attempt to understand the turbulence, and then to introduce self-gravity.  
Introducing all the physical effects at once runs the risk of producing 
results which, while possibly valid, are very difficult to interpret.

The properties of the charged species are based on the data in
\cite{wardle99}.  We make the 
approximation that the dust density is 1\% that of the neutrals.  The
metal ions are assumed to be singly ionised and to have an average mass of 
24\,$m_{\rm p}$ and a mass density of $1.27\times10^{-25}$g\pcm.  The
electron density is then chosen to ensure local charge neutrality.  This
set-up results in an ionisation fraction of approximately
$3\times10^{-7}$ and the system is therefore clearly weakly ionised.

In order to drive turbulence we add velocity increments to the neutral
velocity at each time-step.  The incremental velocity field,
$\delta \bfm{u}$, is defined in a manner similar to \cite{lem09}.  Each 
component of $\delta \bfm{u}$ is generated from a set of waves with wave
numbers, $k = |\bfm{k}|$, $3 \leq k \leq 4$.  The amplitudes of these
waves are drawn from a Gaussian random distribution with mean 1.0 and
deviation 0.33 while the phases of the waves are drawn from a uniform
distribution between 0 and $2 \pi$.  The waves are chosen such that the 
forcing is solenoidal.  It is known that the precise nature of the forcing
will influence the turbulence statistics \cite[e.g.][]{federrath11}. In this 
paper we focus on gaining some insight into the influences of multifluid effects on
turbulence and leave a study of multifluid MHD turbulence under differing driving 
to a later date.

Finally, to ensure a constant energy injection rate and zero net
momentum injection, a quadratic equation must be solved to normalise
$\delta \bfm{u}$.  The resulting normalised incremental velocity field
is then added on to the neutral velocity field.  The rate of injection of energy is the same for
all simulations in this work and is given by $\frac{\dot{E}}{\rho_0 L^2 a^3} = 200$
where $\rho_0$ is the initial (uniform) density, $L$ is the length of side of the computational
domain and $a$ is the (isothermal) sound speed.  In our simulations this energy injection rate
yields steady state turbulence with an RMS Mach number of around 4.5 and an Alfv\'enic Mach 
number of between 3 and 3.5.

Table \ref{table:nomenclature} contains the definition of the nomenclature of the various 
simulations used in this work.

\begin{table}
\centering
\begin{tabular}{lcc} \hline
Name & Resolution ($N_{\rm g}$) & Physics \\ \hline
mf-64 & 64 & Multifluid MHD \\
mf-128 & 128 & Multifluid MHD \\
mf-256 & 256 & Multifluid MHD \\
mhd-256 & 256 & Ideal MHD \\ \hline
\end{tabular}
\caption{Definition of the nomenclature used to refer to the various
simulations presented in this work.}
\label{table:nomenclature}
\end{table}

\section{Analysis}
\label{sec:analysis}

There are two different, but related, aspects to analysing the multifluid simulations presented
here.  The first is associated with understanding the dynamics of the bulk flow, while the second
is focused on the interplay, and differences, between the behaviour of the various species in the
simulation.  In both cases, however, we can use analysis techniques which are broadly similar.

\subsection{Power spectra}
\label{sec:power-analysis}

One of the most popular things to do when faced with a set of results from turbulence simulations
is to take power spectra.  These spectra give information on the relative size of structures
formed by turbulence in the system.  In incompressible hydrodynamic turbulence this is easily
visualised as, for example, the energy residing in vortices of differing sizes.  Standard
Kolmogorov theory predicts that, in the absence of driving or dissipation, the power spectrum will
take on a power-law form, and the exponent of this power law can be predicted in a relatively
straightforward manner.  For incompressible MHD turbulence the situation is a little more
complicated, while for compressible MHD turbulence it becomes more complicated again.  However,
the power spectra of the fluid variables still provide useful physical insight into the type of
turbulence occurring.

The power spectra presented here are calculated by taking a 3D Fourier 
transform of the relevant fluid variable.  The total power residing in a 
spherical shell of inner radius $k$ and outer radius $k + dk$ is then 
calculated and it is this quantity which we present in our power
spectra.  When discussing wave numbers we use the notation $k
\equiv \frac{L}{2 \pi} \tilde{k}$ where $\tilde{k}$ is the usual wave number.  Then $k$ is the number of 
waves which fit into the computational domain for that value of $\tilde{k}$.  In
calculating the power spectra we use $dk = 1$.  The power spectra are taken for different
times and the results presented here are the time-averaged spectra.  For all simulations the spectra
are averaged over $t = [0.2,0.52]$ sound crossing times, or $[0.9, 2.34]$ flow crossing times.

\subsection{The centroid velocity}
\label{sec:centroid-analysis}

The statistics of the centroid velocity are, in principle at least, an observable quantity
and are therefore of some physical interest \citep[e.g.][]{hily09, federrath10}.  Suppose we orient our 
simulation cube such that the $z$ direction is along the line of sight, leaving the $x$ and $y$ 
directions in the plane of the sky.  The centroid velocity is then defined as
\begin{equation}
C(x,y) = \frac{\int \rho(x,y,z) v_z(x,y,z)\,dz}{\int \rho(x,y,z)\,dz}
\label{eqn:centroid-vel}
\end{equation}
The density weighting used here to calculate the average line-of-sight velocity can be thought 
of as some approximation to what happens when observations are made and the average 
line-of-sight velocity is actually the emission-weighted average.

Having defined our centroid velocity we can then consider the centroid velocity increment, 
defined by
\begin{equation}
\delta C_l(\mathbf{r}) = \left<C(\mathbf{r}) - C(\mathbf{r}+\mathbf{l})\right>
\label{eqn:centroid-inc}
\end{equation}
where $\mathbf{r} = (x,y)$ and $\mathbf{l}$ is a vector of length $l$ and the average is 
calculated over all possible directions of $\mathbf{l}$ in the $(x,y)$ plane.

Finally we introduce the associated centroid velocity increment 
structure functions, defined by
\begin{equation}
{\rm CVISF}_p(l) \equiv <|\delta C_l(\mathbf{r})|^p>_{\mathbf{r}}
\label{eqn:cvisf}
\end{equation}
where $p$ is the order of the structure function.  These quantities are discussed in Sect.\
\ref{sec:centroid-analysis}.

\section{Resolution study}
\label{sec:resolution-study}

We performed identical simulations with differing resolutions in order to determine what 
aspects of the results from our highest resolution simulation are well converged.  The 
resolutions used were $N_{\rm g} = 64$, 128 and 256 (see Sect.\ \ref{sec:initial-conditions}).  
Table \ref{table:res-data} contains a summary of our results from the resolution study.

\begin{table}
\centering
\begin{tabular}{cccc} \hline
Simulation & $M_{\rm s}$ & $M_{\rm A}$ & Velocity power 
spectrum \\ \hline
mf-64  & 4.38 & 2.77 & $-2.54 \pm 0.26$ \\
mf-128 & 4.47 & 3.01 & $-2.40 \pm 0.23$ \\
mf-256 & 4.48 & 3.07 & $-2.40 \pm 0.05$ \\
mhd-256 & 4.41 & 3.17 & $-1.81 \pm 0.05$ \\
\end{tabular}
\caption{\label{table:res-data} The time-averaged RMS sonic Mach and
Alfv\'en numbers and the slope of the velocity power spectra for the 
simulations presented in this work.  The velocity power spectrum is fitted 
over the range $[4:N_{\rm g}/16]$ in each case with the exception of mf-64 where the range is
$[4:8]$.  $N_{\rm g}$ is the number of grid zones along a side of our cube. The error quoted is the 
formal error from the least squares fitting.}
\end{table}

\begin{figure}
\centering
\includegraphics[width=8.4cm]{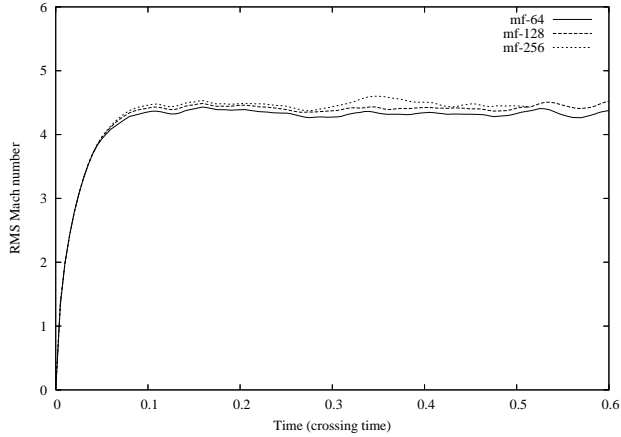}
\caption{Plot of the RMS Mach number as a function of time for each of
the simulations in the resolution study.
} 
\label{fig:res-mach-no}
\end{figure}

Figure \ref{fig:res-mach-no} contains plots of the RMS Mach number as a
function of time for each of the simulations in our resolution study.  It is clear that after an
initial period of time during which the turbulence is building up, each
simulation reaches a quasi-steady state with an RMS turbulent Mach
number of between 4.3 and 4.5.  

The highest Mach number reached occurs, perhaps unsurprisingly, in the $256^3$ simulation.  However, after 
this initial maximum, the average Mach number in each case is rather similar, bearing in 
mind the stochastic nature of the simulations, with the $128^3$ and the 
$256^3$ simulations giving results within about $0.2$\% of each other.  Simulation mf-64 does have a
somewhat lower RMS Mach number (see Table \ref{table:res-data}), indicating that at this low
resolution the global energy dissipation rate is significantly influenced by numerical viscosity.
However, the similarity of the Mach numbers for mf-128 and mf-256 indicate that the global energy 
dissipation rate is well captured with a resolution of $256^3$.

\begin{figure}
\centering
\includegraphics[width=8.4cm]{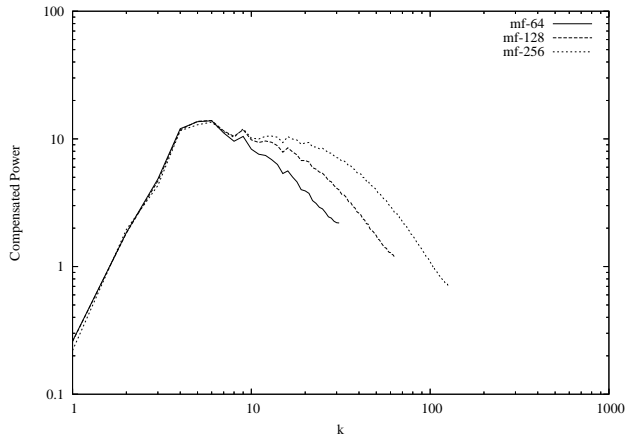}
\caption{Plot of the time-averaged, compensated neutral velocity power 
spectrum for each of the simulations in the resolution study.  See also Table
\ref{table:res-data}.  The spectra are compensated by $k^{2.2}$.
} 
\label{fig:v-spec-res}
\end{figure}

Another measure of whether or not turbulence simulations have converged is the 
slope of the power spectrum of the velocity.  Figure \ref{fig:v-spec-res} 
contains plots of the compensated power spectrum of the neutral velocity for each of the 
three simulations in the resolution study.  While we generally expect this 
power spectrum to behave as a power law, there will be some value of $k$ at 
which the spectrum will steepen due to numerical dissipation effects.  In 
order to get a picture of the range of $k$ over which we can trust our 
results, we have fitted the power spectrum with a power-law over various 
different ranges.  The driving occurs at $3 \leq k \leq 4$ and 
we find that between $k = 4$ and $k \approx \frac{N_{\rm g}}{16}$ ($N_{\rm g}$ being the number of 
grid zones along a side of our cube) our slopes are very similar, keeping in mind the formal errors, 
across all resolutions (see Table \ref{table:res-data}).  These results give support to
the idea that these simulations have resolved the turbulent cascade, at least up to a value of $k$ of 
around $\frac{N_{\rm g}}{16}$ zones.  Even so, when interpreting the results in a quantitative
fashion, the size of the formal errors must be kept in mind and the best way to reduce these is
through using higher resolution simulations.  The results of \cite{federrath10} suggest that
caution must be used when interpreting power spectra at high $k$.  In their particular work
where FLASH3 was used they found the power spectra to be susceptible to numerical dissipation
below length scales of about 30 grid zones.  Our resolution study here demonstrates that the
equivalent length scale, for velocity power spectra, for HYDRA is about 10 -- 15 grid zones.

\section{Results}
\label{sec:results}

Following the results of our resolution study we are reasonably confident that 
our $N_{\rm g} = 256$ simulation is well enough resolved to draw some useful 
physical conclusions about the behaviour of driven, multifluid MHD turbulence.

\begin{figure}
\centering
\includegraphics[width=8.4cm]{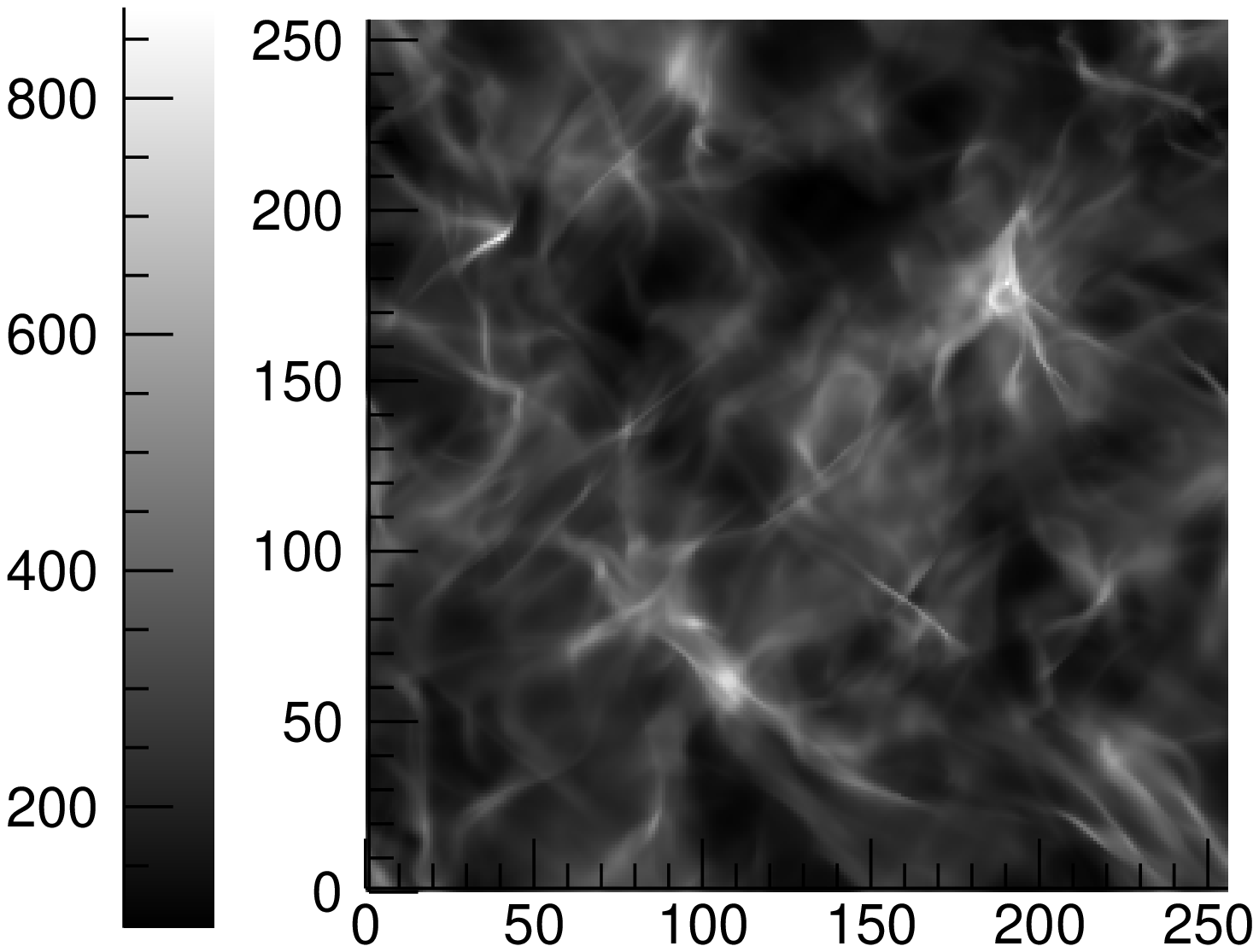}
\includegraphics[width=8.4cm]{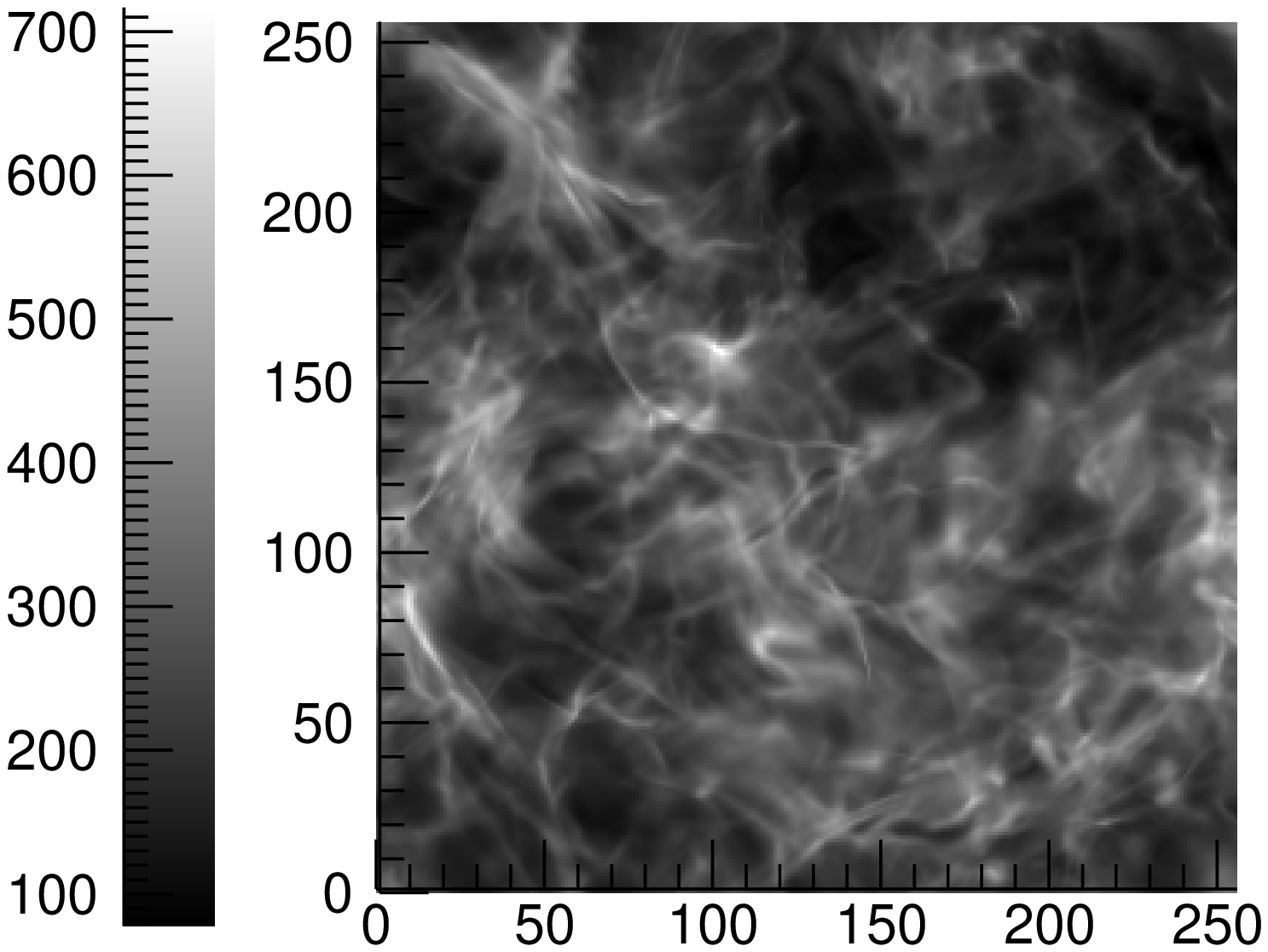}
\caption{Distribution of the column density for mf-256 (upper panel) and mhd-256 (lower panel) at time 
$t=0.5$\,$t_{\rm c}$, assuming the line of sight is along the $z$ axis.  The units on the $x$
and $y$ axes are grid zones and the density is in units of $10^4$\pcm.
} 
\label{fig:col-density}
\end{figure}

Figure \ref{fig:col-density} contains plots of the (neutral) column density for the multifluid
and ideal MHD simulations.  It is apparent that there is somewhat more small scale structure in mhd-256 than in
mf-256.  The maxima and minima of the column density are broadly similar in each case.

\subsection{Energy dissipation}
\label{sec:dissipation}

\begin{figure}
\centering
\includegraphics[width=8.4cm]{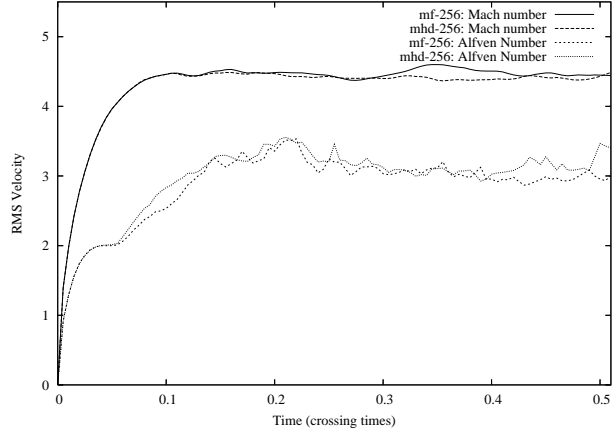}
\caption{Plot of the RMS Mach and Alfv\'en numbers as functions of time for mf-256 and mhd-256.}
\label{fig:mf-mhd-rms}
\end{figure}

Figure \ref{fig:mf-mhd-rms} contains plots of the behaviour of the mean, 
mass-weighted RMS Mach and Alfv\'en numbers \citep[see][]{lem09, dos11}.  It is clear
that the Mach number reached in each case is very similar.  Table
\ref{table:res-data} demonstrates this in quantitative terms with the
difference in time-averaged RMS Mach number after $t=0.2$\,$t_{\rm c}$ being around than 1.6\%.  
Therefore, at these length scales and with this numerical resolution, we do not detect any difference 
in the energy dissipation rate due to multifluid effects.  Note that the RMS Mach number is actually
a simple measure of the RMS velocity on the computational grid as the simulations are isothermal.  On
the other hand, the RMS Alfv\'en number is a more complicated quantity, given that the Alfv\'en speed
is a spatially and temporally varying quantity.  It is clear that the difference in the Alfv\'en
numbers between mhd-256 and mf-256 are rather small, with a difference of only around 3\%.  This is
quite interesting, given that ambipolar diffusion does diffuse magnetic energy and hence would be
expected to lower the Alfv\'en speed generally.  Hence, taken together, the similarities between
mf-256 and mhd-256 with respect to their Mach and Alfv\'en numbers indicates that the Alfv\'en speed
is not, in fact, lowered significantly on a global scale by multifluid effects.

The results presented in Papers I and II imply that the energy dissipation rate is higher in multifluid 
MHD systems.  We would expect, then, that the mean Mach number would be lower in mf-256 than in
mhd-256.  However, it is clear from Table \ref{table:res-data} that this is not the case.  If,
instead of the time-averaged mass-weighted RMS Mach number, we calculate the time-averaged RMS
Mach number without mass-weighting, we find that the RMS Mach numbers are 4.18 and 4.10 for
mhd-256 and mf-256 respectively.  This is a little more in line with what we would naively expect although, on
the face of it, it still does not indicate significantly increased energy dissipation due to multifluid 
effects.  When considering this issue, it is significant that the main contribution to the RMS Mach numbers in each
of mhd-256 and mf-256 comes from large scale motions - i.e.\ motions close to the driving scale.  Since the multifluid
effects operate at relatively short length scales it, perhaps, is not too surprising that the RMS Mach number is not
effected much by the inclusion of these effects.  When reconciling this result with those of Papers I and II it is worth
recalling that these latter results were calculated for decaying turbulence.  As turbulence decays energy initially
injected at large scales cascades to ever smaller scales.  Thus decaying multifluid turbulence will exhibit a faster Mach
number decay as, in contrast to driven turbulence, the main contribution to the measured Mach number will come from ever smaller 
scales, and thus be more influenced by multifluid effects.

\begin{figure}
\centering
\includegraphics[width=8.4cm]{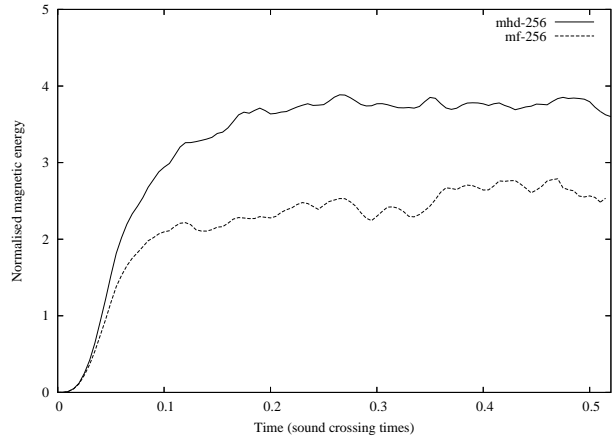}
\caption{Plot of the total perturbed magnetic energy in the computational domain as a function of time for 
mf-256 and mhd-256.}
\label{fig:mf-mhd-mag}
\end{figure}

There is a further effect to take into account.  Closer examination of the data from mf-256 and mhd-256 reveals that, 
while the kinetic energies in each simulation are very similar, the magnetic energy in the multifluid simulation is much 
decreased (see Figure \ref{fig:mf-mhd-mag}).  This reduction in the magnetic energy is due to the presence of ambipolar 
diffusion and shows that the multifluid effects do have a significant impact on the energetics of the
system.  The fact that the RMS Mach number is not strongly influenced by multifluid effects is partly a
reflection of the fact that the energy in the system is dominated by kinetic energy.  Some of this
kinetic energy is converted into magnetic energy but, even in mhd-256, this is a relatively small
proportion: on average the magnetic energy is around 40\% of the kinetic energy.  The fact
that this converted energy is then lost faster in mf-256 has relatively little impact on the large
reservoir of kinetic energy, and therefore has a minor influence on the RMS Mach number.  On a side note, it is
interesting to note the results of \cite{federrath11} in which studies of dynamo action in MHD simulations of
turbulence.  These authors found that, as might be expected, solenoidally forced turbulence induces stronger dynamo
action.  The relevant saturation levels for the ratio of magnetic to kinetic energy from that study are around 0.03,
considerably smaller than the results presented here.  However, the significant differences in terms of initial
conditions between this latter study and the one presented here make it difficult to draw any firm conclusions about the
origins of the differences between the results.

In the case of decaying turbulence, the magnetic energy becomes approximately equal to the kinetic
energy as the decay process continues (see Figure 6 of Paper I) and hence the multifluid effects can
have a more significant impact on the overall energetics of the system, resulting in a faster decay
rate.

\subsection{Power spectra}
\label{sec:power-spectra}

We now turn to the power spectra of the various quantities of interest
in these simulations.  The nature of these spectra will give us some idea of the influence
of the multifluid effects on the turbulence.  As noted previously, we time-average all power spectra
discussed here over the time-interval $[0.2\,t_{\rm c}:0.52\,t_{\rm c}]$. Table 
\ref{table:spectra-exp} contains the exponents of the power spectra, fitted assuming the spectrum goes 
as $k^{\alpha}$.  The fitting is carried out over the same range as that in Table 
\ref{table:res-data} and so we are reasonably confident that the exponents we find are not significantly 
influenced by numerical resolution effects.

\begin{table*}
\centering
\caption{Exponents of the power spectra of the various variables in the
mf-256 and mhd-256 simulations.  The spectra were fitted as power
laws in the range $4 \leq k \leq \frac{N_{\rm g}}{16}$.
\label{table:spectra-exp}}
\begin{tabular}{lccccc} \hline
 & mhd-256 & \multicolumn{4}{c}{mf-256} \\
 & & Neutrals & Electrons & Dust & Ions \\ \hline
Density & -0.60 & -0.83 & -0.99 & -1.11 & -0.99 \\
Velocity & -1.81 & -2.40 & -2.17 & -2.19 & -2.17 \\
$\mathbf{B}$-field & -1.73 & \multicolumn{4}{c}{-2.26} \\ \hline
\end{tabular}
\end{table*}

\subsubsection{Velocity power spectra}

\begin{figure}
\centering
\includegraphics[width=8.4cm]{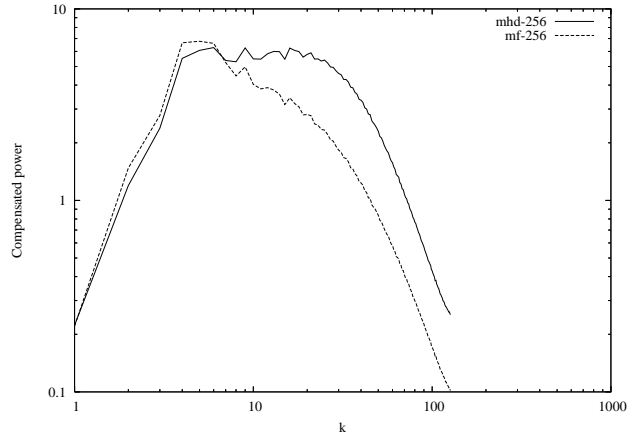}
\caption{Time-averaged compensated, spherically integrated velocity power spectra for the 
neutrals in mf-256 and the bulk flow in mhd-256.  The spectra are compensated with $k^{1.8}$.
} 
\label{fig:mf-mhd-v-power}
\end{figure}

The compensated velocity power spectra for the neutrals in mf-256 and the bulk fluid in mhd-256 
are shown in figure \ref{fig:mf-mhd-v-power}.  The power is compensated with $k^{1.8}$ which is
the measured exponent of the velocity spectrum of mhd-256 (see Table
\ref{table:spectra-exp}).  It is clear, as indicated also in Table \ref{table:spectra-exp},
that the multifluid effects modify the velocity power spectra of the neutrals
significantly, even on rather large length scales.

\begin{figure}
\centering
\includegraphics[width=8.4cm]{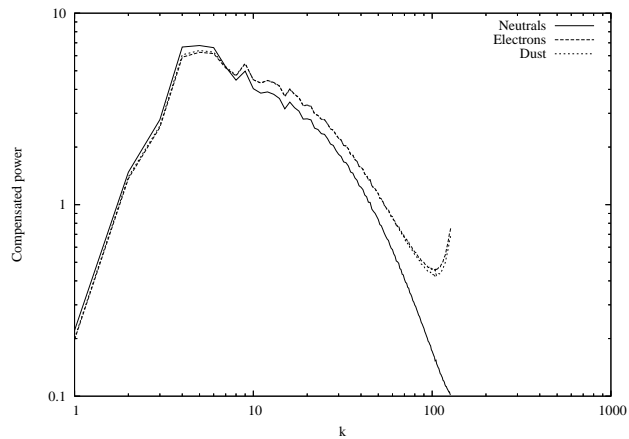}
\caption{Time-averaged, compensated, spherically integrated velocity power spectra for the 
neutrals, electrons and dust in mf-256.  The spectrum for ions is virtually identical to that for
electrons and is omitted.
\label{fig:n-elec-v-power}
} 
\end{figure}

Figure \ref{fig:n-elec-v-power} contains plots of the spherically integrated, compensated
velocity power spectra for the neutrals, electrons and dust in mf-256.  We have omitted a plot of
the spectrum for the ions as it is virtually identical to that for the
electrons.  It is clear that the velocity power spectra for the neutrals and
charged species do differ to some extent on these length scales (i.e.\ all scales up to 0.05\,pc).  
The spectrum for the charged species is somewhat harder than the neutrals.  This is in line with what would 
be expected in a multifluid system affected by ambipolar diffusion: as ambipolar diffusion becomes more
important the neutrals behave more like a hydrodynamic system which characteristically has
a softer velocity power spectrum than an MHD system.  The charged species remain more
affected by the magnetic field and therefore retain more strongly the characteristics of MHD
turbulence.  Another way of looking at this is to observe that, from the data in Table
\ref{table:spectra-exp}, the power spectra of the velocity for the charged species have exponents
which lie between that for the neutrals and that for ideal MHD.  Again, this is what would be
expected given the more direct link between the behaviour of the charged species and that of the
magnetic field.  At very high $k$ there is a slight difference between the dust and electron power 
spectrum, almost certainly due to the Hall term in the induction equation.  However, we should refrain
from attaching quantitative physical significance to this as the length scales involved at these values of 
$k$ are too short to be reliably resolved in the simulation.

\subsubsection{Density power spectra}

\begin{figure}
\centering
\includegraphics[width=8.4cm]{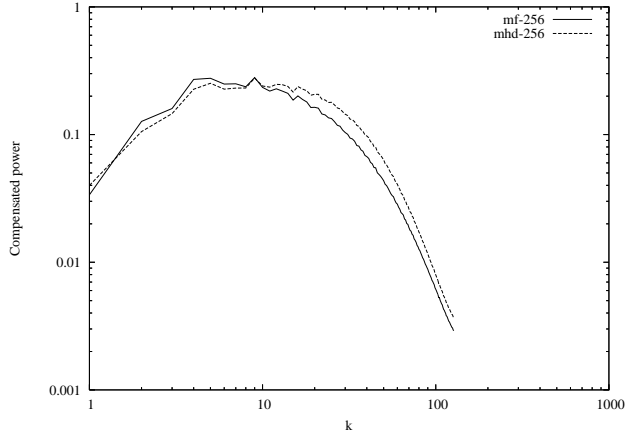}
\caption{Plots of the time-averaged power spectra for the neutral density in mf-256 and
the density in mhd-256.  The spectra are compensated using $k^{0.6}$ (see Table
\ref{table:spectra-exp}).}
\label{fig:rho-power-mf-mhd}
\end{figure}

Density power spectra, while not as dynamically relevant in supersonic turbulence as
velocity power spectra, nonetheless provide an insight into the size of structures formed by
the turbulence \citep[e.g.][]{ossenkopf02}.  This can also be investigated in a different way using the probability density 
function (PDF) of the density which we discuss later (Sect.\ \ref{sec:pdf}).  Figure 
\ref{fig:rho-power-mf-mhd} contains plots of the time-averaged (neutral) density power spectra 
for mf-256 and mhd-256.

There is less power in the bulk density of mf-256 than mhd-256 for all $k \geq 10$ as one might
expect, given the similarity between the multifluid effects and viscosity and also given the
behaviour of the velocity power spectra.  Interestingly, there is marginally less power in mhd-256 than 
mf-256 at $k < 10$. This difference is rather small and is likely to be due to diffusive effects causing
an inverse cascade, as would be expected since they smooth out small scale structures into larger scale
ones.

\begin{figure}
\centering
\includegraphics[width=8.4cm]{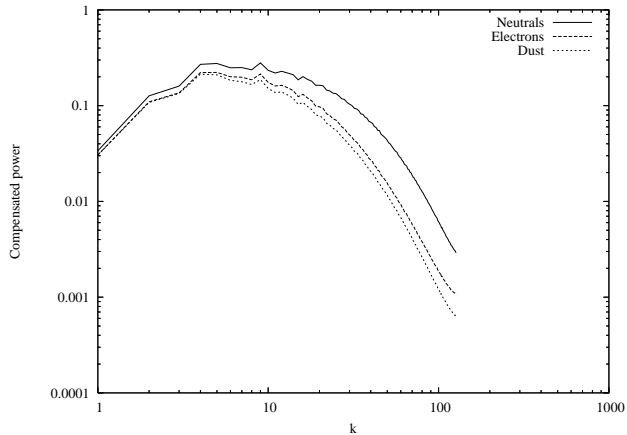}
\caption{Plots of the the time-averaged, normalised power spectra of the density for all species in 
mf-256.  The spectra are normalised and compensated by $k^{0.6}$ for ease of comparison (see Table 
\ref{table:spectra-exp}).  The power spectrum for the ions is not plotted as it is identical
to that for electrons.}
	\label{fig:rho-all-power-mf}
\end{figure}

\begin{figure}
\centering
\includegraphics[width=8.4cm]{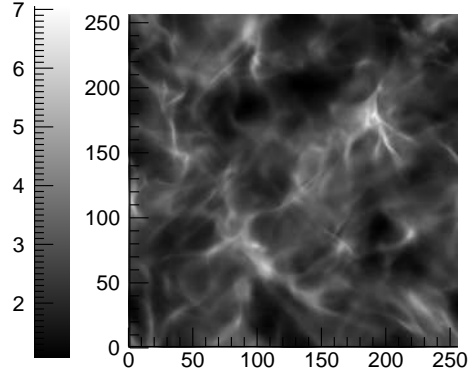}
\caption{Distribution of the column density for the dust in our mf-256
simulation at time $t=0.5$\,$t_{\rm c}$, assuming the line of sight is along 
the $z$ axis.  The units on the $x$ and $y$ axes are grid zones and the 
density is in units normalised by the initial neutral density.
Comparison with the upper panel in figure \ref{fig:col-density}
indicates there is less small-scale structure in the dust distribution.
} 
\label{fig:dust-col-density}
\end{figure}

While there is somewhat less power in the density variations in mf-256 than mhd-256, there are 
striking differences between the power in the density variations of the neutral and charged
species in mf-256 (see Figure \ref{fig:rho-all-power-mf}).  Hence we have a situation where the neutral 
power spectrum is more qualitatively similar to that obtained from ideal MHD, while the charged species 
(which are more directly influenced by the magnetic field) have significantly less power at virtually 
all length scales.  A careful comparison of figure \ref{fig:dust-col-density} and the upper panel of 
figure \ref{fig:col-density} reveals that, even by eye, there is less structure in, for example, the 
dust distribution than the neutral distribution.

The inclusion of multifluid effects leads to the introduction of terms in the
induction equation which are diffusive so it is not a surprise that there is less structure
in a multifluid MHD system than an ideal MHD one.  It is also, perhaps,
not surprising that this lack of structure is particularly marked in the charged species
which are more directly affected by the dynamics of the magnetic field.

\subsubsection{Magnetic field power spectra}
\label{sec:b-pow}

\begin{figure}
\centering
\includegraphics[width=8.4cm]{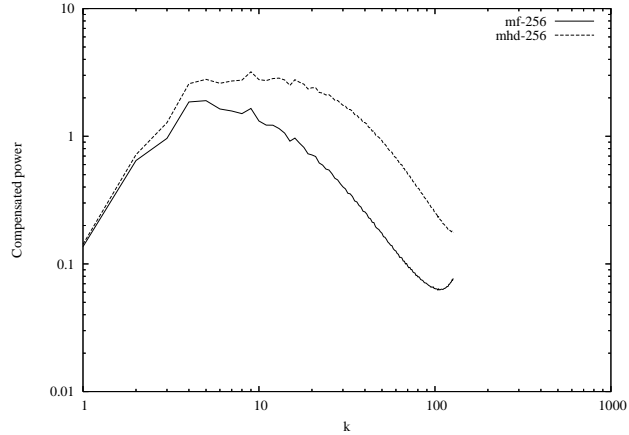}
\caption{Time-averaged, compensated, spherically integrated magnetic field power spectra
for mf-256 and mhd-256.  Each power spectrum is compensated with $k^{7/4}$.} 
\label{fig:b-pow}
\end{figure}

We now turn our attention to the power spectra of the magnetic fields in each simulation
(see Fig.\ \ref{fig:b-pow}).  In keeping with the velocity and density power spectra
(Figs.\ \ref{fig:rho-power-mf-mhd} and \ref{fig:mf-mhd-v-power}), there is less structure
on all scales, even at the driving scale, in the multifluid system than in 
the ideal MHD system.  This is due to the presence of ambipolar diffusion which, being a 
standard diffusion term in the induction equation, will smear out any small scale structure.  
This result implies that multifluid effects can have an impact on the nature of molecular cloud 
turbulence even at fairly large length scales - much larger than those inferred from, for
example, the observations of \citet{li08, li10}.  We address this apparent contradiction in a forthcoming 
paper.

It is worth noting again that the Hall effect is present in this system, albeit at a much lower
level than ambipolar diffusion.  As in Paper I and Paper II we find little evidence of the
Hall effect at length scales which are numerically resolved in our power spectra.  At very short
length-scales ($k \geq 100$) we can see an upturn in the power spectrum for the magnetic field and
the charged species (Figure \ref{fig:n-elec-v-power}) and this is almost certainly the influence of the 
Hall effect where, as expected, we see a decoupling at short length scales between the neutrals
and the charged species, as well as the magnetic field.  The phenomenon of the Hall effect having
somewhat less impact than would be expected was also discussed in \cite{jd12} who deduced that the 
presence of ambipolar diffusion can effectively short circuit the Hall current, thereby inhibiting the 
Hall effect even in systems with significant Hall resistivity.

\subsection{Probability Density Functions}
\label{sec:pdf}

Probability density functions (PDFs), in particular of the mass density, are particularly
interesting when considering the influence molecular cloud turbulence may have on star
formation.  It has long been known that isothermal, supersonic turbulence results in a log-normal
PDF of the density \cite[e.g.][]{vaz94, pas98}.  This is due simply to the fact that the density
of a particular fluid element is a product of its initial value and all of the enhancements at
each of the shocks it has encountered.  Since the shock strengths are randomly distributed, the
density is thus the result of a product of random variables and hence its distribution is
log-normal.  Several researchers have shown that the addition of gravity to the physics of the system 
adds a long, high mass tail to the PDF \cite[e.g.][]{klessen00, dib05}, presumably due to the influence 
of gravity which will produce accumulations of mass in small regions which would not be expected to 
happen to the same extent with isothermal turbulence.  This would lead to a positive skewness of
the density distribution or, to say it another way, the distribution being skewed right.

\begin{figure}
\centering
\includegraphics[width=8.4cm]{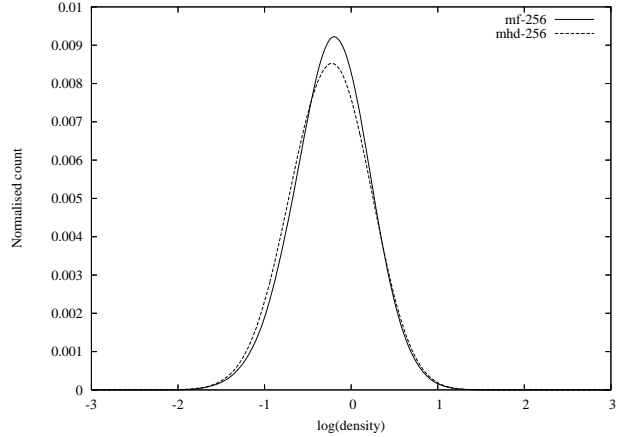}
\caption{Time-averaged PDF of the normalised bulk density for mf-256 and mhd-256.} 
\label{fig:neutral-pdf}
\end{figure}

Figure \ref{fig:neutral-pdf} contains plots of the PDF of the logarithm of the density of
the bulk flow of mhd-256 and mf-256.  Table \ref{table:pdf-stats} details the standard deviation, skewness and 
excess kurtosis of the distributions of all of the densities in our simulations.  Clearly all of the densities 
in each of mf-256 and mhd-256 have very close to a log-normal distribution.  The
logarithm of the bulk mass density is very slightly platykurtic in both mf-256 and
mhd-256, while the distributions of the charged densities are a mix of leptokurtic and
platykurtic.  However, these differences in the kurtosis are rather minor.  The skewness of these 
distributions is also small, indicating again how close to log-normal the PDF of the density of the 
bulk flow is.

\begin{figure}
\centering
\includegraphics[width=8.4cm]{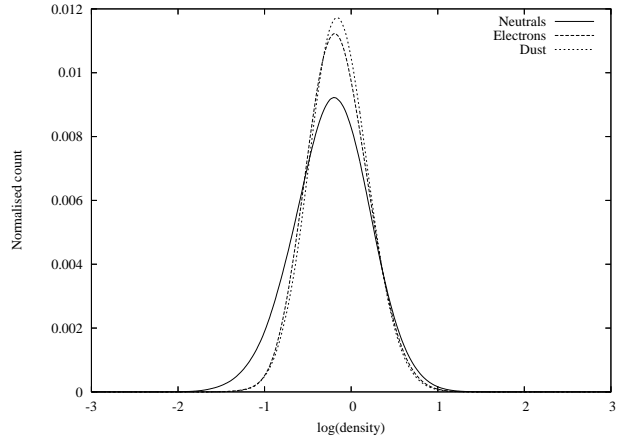}
\caption{Time-averaged PDF of the normalised densities of the neutrals, electrons and dust in 
mf-256.  The ion density PDF is not shown as it is identical to that of the electrons.} 
\label{fig:neutral-electron-pdf}
\end{figure}

Examining Figure \ref{fig:neutral-electron-pdf} and Table \ref{table:pdf-stats} we see that the charged species are slightly 
less platykurtic (flattened) than the neutrals in mf-256.  We can understand this as follows: the 
magnetic field has, of course, a pressure associated with it which will resist both compression and 
rarefaction.  Since the charged species are, by definition, more closely tied to the magnetic field 
than the neutrals we expect that the neutrals will undergo more severe compressions and rarefactions 
than the charged species.  Thus the distribution of the charged species densities' should be 
more peaked around their mean than the neutrals. The standard deviation
of our neutral species implies a value for the constant of proportionality
between the Mach number and standard deviation \citep[the ``$b$'' parameter in, for
example,][]{federrath08} of 0.31, in keeping with the solenoidal forcing
used in the work presented here.  That the charged species have a higher
standard deviation is unlikely to be of {\em direct} importance to models of
the Initial Mass Function as the charged species are insignificant in terms
of their mass, and hence their direct contribution to gravitational collapse.  
That the charged species are less abundant in high and low density regions
than in intermediate density regions does have implications for the behaviour
of the resistivities in these regions and hence for the influence of ambipolar
diffusion and fragmentation.  A detailed study of this is, however, beyond the
scope of this paper.

\begin{table*}
\centering
\caption{Statistics of the probability distribution functions of the logarithm (to the base 10) of the 
	various densities in the simulations presented here.
\label{table:pdf-stats}}
\begin{tabular}{lccccc} \hline
 & mhd-256 & \multicolumn{4}{c}{mf-256} \\
 & & Neutrals & Electrons & Dust & Ions \\ \hline
Standard deviation & 0.46 & 0.44 & 0.36 & 0.34 & 0.36 \\
Skewness & -0.050 & -0.12 & 0.15 & 0.082 & 0.15 \\
Kurtosis & -0.11 & -0.071 & -0.005 & 0.057 & -0.005 \\ \hline
\end{tabular}
\end{table*}

\subsection{Centroid velocity statistics}
\label{sec:centroid-velocity}

\begin{figure}
\centering
\includegraphics[width=8.4cm]{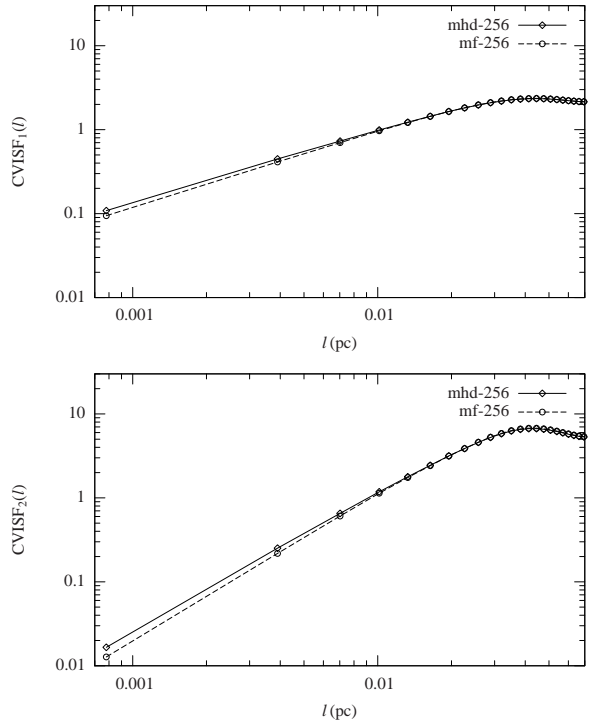}
\caption{Plots of the first order (top panel) and second order (bottom panel) structure functions of the 
centroid velocity increment for the neutral species in mf-256 and the bulk flow in mhd-256.  Both 
these clearly follow power laws in $l$ up to lengths of around one quarter of the computational domain. \label{fig:cvisf}} 
\end{figure}

We now turn to the statistics associated with the centroid velocity (see Sect.\
\ref{sec:centroid-analysis}).  Figure \ref{fig:cvisf} contains plots of the first 2 structure
functions of the centroid velocity (equation \ref{eqn:cvisf}).  Table \ref{table:cvisf} contains
the absolute and relative exponents of these structure functions found by the fit CVISF$_p(l)
\propto l^{\zeta_p}$.  From both Figure \ref{fig:cvisf} and Table \ref{table:cvisf} it is clear that the 
introduction of multifluid effects appears not to influence the centroid velocity structure functions of 
the neutral flow significantly.  

It is worth noting that the absolute exponents of the structure functions differ somewhat from those given in
\cite{federrath10} with a difference of around 15\% for the $p=1$ structure function.  We performed a further ideal MHD 
simulation with a resolution of $512^3$ to assess whether resolution issues may be affecting our results in this regard and we
were able to confirm our results for the exponents.  Since the simulations of \cite{federrath10} were hydrodynamic we
conclude that the reason for the differences in our results is due to the presence of dynamically significant magnetic
fields.

\begin{table}
\centering
\begin{tabular}{lcc}\hline
Simulation & mf-256 & mhd-256 \\ \hline
$\zeta_1$ & 0.32 (0.71) & 0.32 (0.69) \\
$\zeta_2$ & 0.65 (1.43) & 0.65 (1.39) \\
$\zeta_3$ & 1.00 (2.19) & 1.00 (2.14) \\ \hline
\end{tabular}
\caption{\label{table:cvisf} The exponents of the CVISF$_p(l)$, normalised to $p=3$ for $p=1,2,3$.
The numbers in brackets are the absolute exponents.}
\end{table}

\section{Conclusions}
\label{sec:conclusions}

We have presented the results of 4-fluid, MHD turbulence in molecular clouds.  The parameters
chosen were matched to the observations of molecular clouds reported in \citet{cru99}.  We
performed a resolution study to ensure that the results presented, in terms of both the power
spectra and the RMS Mach number, were not strongly influenced by resolution effects.  

We found that the RMS mass-weighted Mach number was not significantly decreased with the inclusion of multifluid effects, 
in spite of multifluid effects providing an effective pathway for removal of magnetic energy.  The overall energy dissipation, 
then, appears not to be significantly affected by multifluid effects.  This phenomenon is a result of the fact that the
majority of the energy in a driven, turbulent system with the properties of a molecular cloud resides in kinetic energy and so 
an enhanced removal of magnetic energy does not significantly effect the global energetics of the system.

We found that both the velocity power spectra of the neutrals and the magnetic field power spectra
were strongly influenced by multifluid effects at all length-scales up to the driving scale
(0.05\,pc).  In multifluid MHD, the velocity power spectrum of the neutrals is found to have
considerably less structure at all length-scales than those of the charged species.  The electron
and ion fluids behave almost identically to each other, while the dust displays some slight
differences with the other charged species.  This latter point arises largely from the requirement
of local charge neutrality and the fact that the electrons and ions are the dominant charge
carriers in the system.  We also find, at very short length-scales, some signs of the Hall effect in
the power spectra of the charged species.

Thus we find, in common with the results presented in Papers I and II that multifluid 
effects are important up to quite large length-scales.  This result appears to be in conflict with
the observations of \cite{lh08} and we deal with this apparent contradiction in a forthcoming
paper.

The differences in the power spectra of the bulk densities in the multifluid and ideal MHD systems
are not as dramatic as those in the other variables, although there is somewhat less power in the
distribution of the neutral density in the multifluid simulation than the bulk density in the ideal
MHD one.  Interestingly, there is {\em less} power in the spectrum of the charged species density
than the neutral species in the multifluid system.  This is the opposite of what is found for the
velocity power spectra.

For all species, and for all cases, the density PDFs are found to be very close to log-normal.  The
neutral species in the multifluid simulation has a slightly less platykurtic (flattened)
distribution than that in the ideal MHD case.  The charged densities in the multifluid simulation have, 
in turn, less platykurtic distributions than the neutrals in that simulation with the dust 
actually having a leptokurtic distribution.  Each of these results is explained in terms of the 
ambipolar diffusion experienced by the multifluid system, and the fact that the charged species are 
more directly influenced by the magnetic field than the neutrals.

Finally, we find that multifluid effects do not significantly influence the centroid velocity increment structure
functions.

\section*{Acknowledgements}
The computations for this work were carried out as part of the DEISA-DECI 6 project ``D-Sturb''.  We 
also acknowledge the results presented in this paper were partly achieved using the PRACE Research 
Infrastructure resource JUGENE based in Germany at the J\"ulich Supercomputing Centre.

\label{lastpage}

\end{document}